# An Efficient Coding Method for Coding Region-of-Interest Locations in AVS2

Mingliang Chen[1], Weiyao Lin[1*], Xiaozhen Zheng[2]
[1]Department of Electronic Engineering, Shanghai Jiao Tong University, China (* Corresponding Author)
[2] Research Department of Hisilicon Semiconductor and Component Business Department, Huawei Technologies, China

## Abstract

Region-of-Interest (ROI) location information in videos has many practical usages in video coding field, such as video content analysis and user experience improvement. Although ROI-based coding has been studied widely by many researchers to improve coding efficiency for video contents, the ROI location information itself is seldom coded in video bitstream. In this paper, we will introduce our proposed ROI location coding tool which has been adopted in surveillance profile of AVS2 video coding standard (surveillance profile). Our tool includes three schemes: direct-coding scheme, differential- coding scheme, and reconstructed-coding scheme. We will illustrate the details of these schemes, and perform analysis of their advantages and disadvantages, respectively.

## I. Introduction and Related Work

Region-of-Interest (ROI) location information for video sequence is of increasing importance in many applications. For example, ROI location information can help encoder decide quantization parameters to enhance the resolution of the interest regions. Specifically, in video conference applications, identified ROIs can be used to guarantee the video qualities of the interested region in order to improving the user experiences.

Various ROI-based coding methods have been developed. Chai. D et al. [1] extract face region using skin-color map to obtain ROI location. Chen et al. [2] use robust skin-color detection to obtain ROI position information for the coding. Menser et al. [3] analyze the video contents by face detection and tracking technology, and then apply them to the application of ROI coding. Hu et al. [4] categories macroblocks inside one frame into three different region types and then perform rate control by allocating more bits to those region types that more interesting information. However, these methods all treat ROI location information as a reference to decide coding parameters, e.g. quantization parameter, for macroblocks, rather than a part of video contents which can be encoded into the bitstream. In our analysis, ROI location information contains lots of contents relevant to the corresponding video. It can be used as a efficient way that describes the video contents' characteristic. Moreover, ROI location information records trajectory of the target objects, which is especially usefully in surveillance videos. Therefore, it is reasonable to encode ROI location information into the video bitstream.

In the following parts of this paper, we will describe our proposed ROI location coding tool which has been adopted in surveillance profile of AVS2 video coding standard. We will illustrate three coding schemes, which are direct-coding scheme, temporal differential-coding scheme, and reconstructed- coding scheme. Besides, the analysis and comparison of these schemes will also be presented.

The rest of the paper is organized as follows: Section II describes the details of the three schemes in our ROI coding method. Section III shows the experiment results and analyzes the coding performances of these schemes. The conclusions are drawn in Section IV.

## II. Details of Our ROI Coding Tool

### *A. Definitions*

In our design, we assume ROIs in videos can be bounded by rectangles and the locations of ROIs can be described and coded by the corner points of these bounding rectangles. According to this assumption, we define ROI location information as:

$$R_i = \{label_i, x_i, y_i, w_i, h_i\} \quad (1)$$

where $R_i$ indicates the $i$-th ROI in one frame, $label_i$ is the index of $R_i$ indicating the ROIs with the same properties, e.g. the same object: a person, a car, etc., in the whole video sequences, $x_i$ and $y_i$ are the horizontal and vertical coordinates of the top-left corner pixel of $R_i$, and $w_i$ and $h_i$ are the width and height of the $R_i$, respectively. With these five parameters, we can uniquely determine the location and the property of one ROI in a frame. Fig. 1 illustrates the parameters in Eqn. (1).

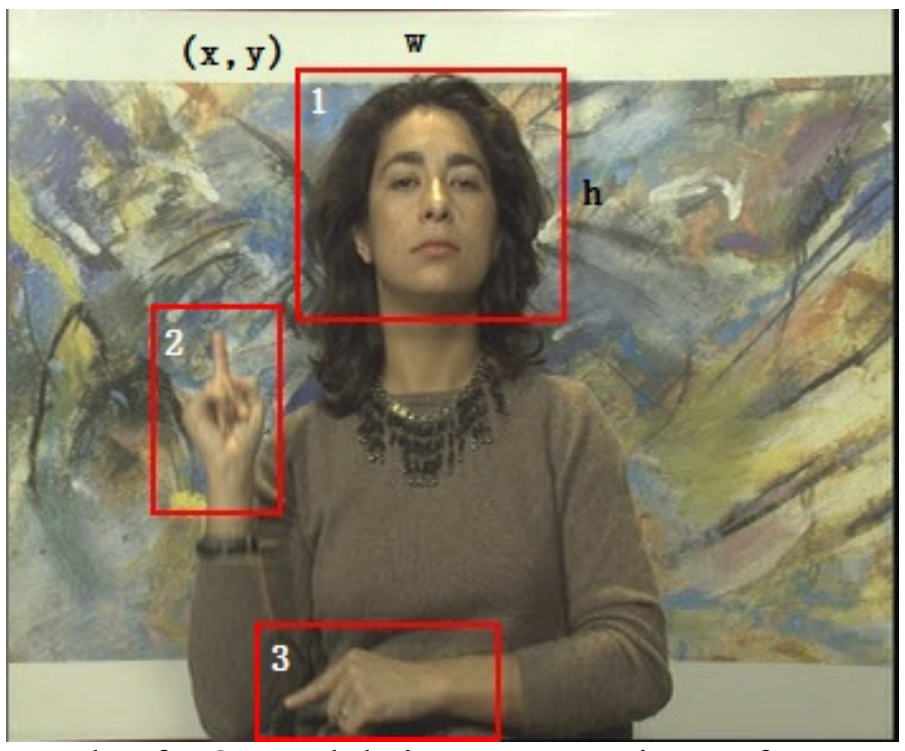

Fig. 1 An example of ROIs and their parameters in one frame

Then, we define the differentiation of two corresponding ROIs as:

$$D = R - R' = \{x - x', y - y', w - w', h - h'\} \quad (2)$$

where $R$ and $R'$ are two corresponding ROIs in adjacent encoded frames. It will be applied in differential-coding scheme and extension scheme.

Before describing the schemes in detail, three important points need to be mentioned:

1) ROIs in videos can either be selected manually or be detected by some state-of-art detection techniques [5].

2) The order of ROIs in a frame is ordered by their labels indices. In the proposed method, we allocate ROI label indices according to the order of their appearing time.

3) The ROI location information will be added to head of each frame's bitstream.

In the following, we will describe the three proposed schemes, respectively.

### B. Direct-Coding Scheme

In the direct-coding scheme, we encode the raw ROI location information directly without any compression.

During the procedure, we process the information in each frame independently, i.e. we can recover the ROI location information in one frame without any additional information of other frames. More specifically, we first encode the label index difference between the current ROI and the previous one in a frame (if it is the first ROI in the ROI sequence in one frame, the difference will be the same as the label number -1). Then, we encode the four parameters of an ROI (Eqn. (1)) directly into the bitstream. We encode the ROIs by the above method until the end of the ROI sequence in one frame. And the above process will be repeated for each frame.

The detailed process of the direct-coding scheme is shown by Algorithm 1 and Fig. 2 shows an example of using the direct-coding scheme to encode ROIs in a frame.

---

**Algorithm 1:** Direct-coding scheme

**Input:** ROI location information in one frame (assume that there are totally $n$ ROIs)
**Output:** Bitstream containing ROI information

encode $n$ into the bitstream;
**for** $i$ = 1 to n
  $R_i$ = the $i$-th ROI in the sequence,
    with parameters $\{label_i, x_i, y_i, w_i, h_i\}$;
  **if** ($i$ == 1)
    $d_i = label_i - 1$;  /*label index difference between the current ROI */
        /*and the previous ROI in the current frame*/
  **else**
    $d_i = label_i - label_{i-1} - 1$;
  **end**
  encode $d_i, x_i, y_i, w_i, h_i$ into the bitstream sequentially;
**end**

---

We can see that this scheme is simple and easy to operate. It is suitable to encode the ROI location information when a few of ROIs need to be encoded in the ROI sequence per frame.

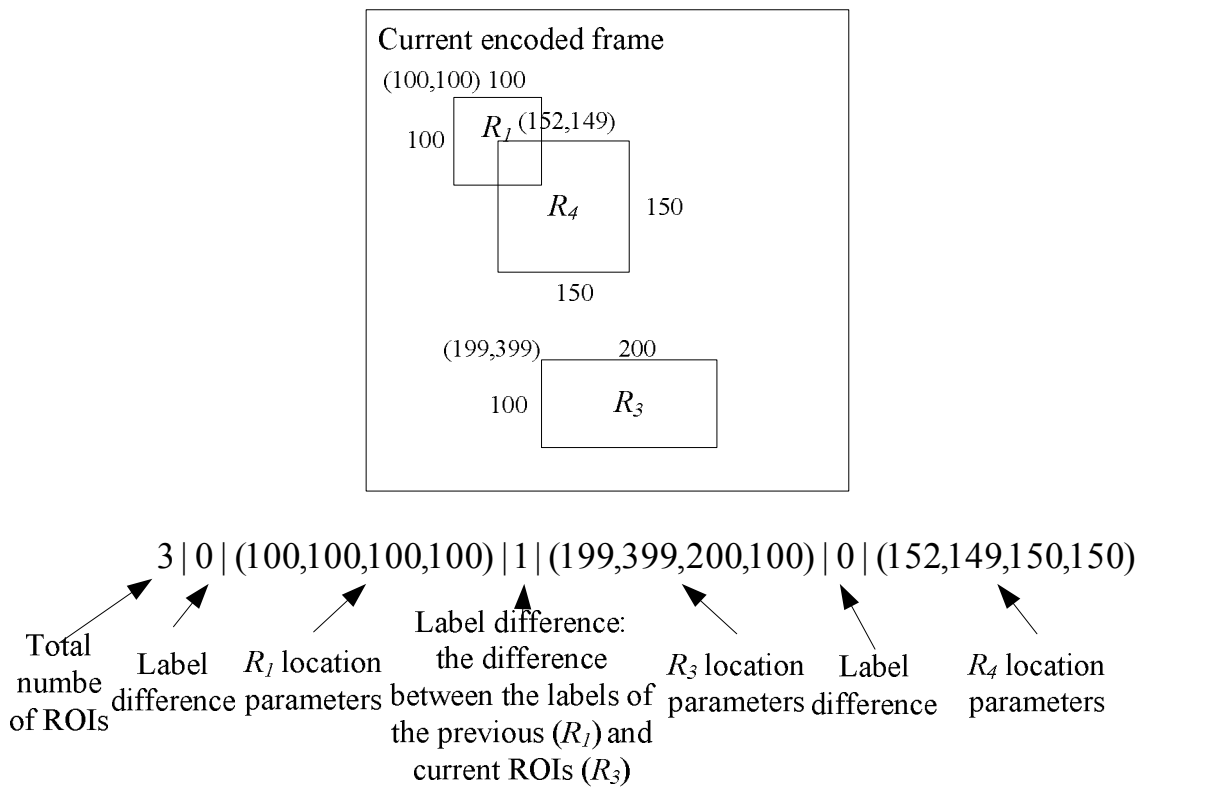

Fig. 2 An coding example of using the direct-coding scheme: assume that the label of $R_i$ is $Label_i$=$i$.

### C. Differential-Coding Scheme

Although direct-coding scheme can encode ROI location information into the bitstream, when large quantities of ROIs are required to be encoded, the bitstream of the ROI location information will become huge and non-negligible, and will add a burden to the video bitstream. Therefore, we developed another scheme, named as differential-coding scheme, to further improve the ROI coding efficiency.

In this scheme, we encode the ROI location information with the temporal differential coding idea to reduce the coding redundancy. Since temporally neighboring ROIs has high correlation, by encoding the temporal difference of the corresponding ROI location parameters of one ROI, redundancies can be reduced. The detailed process of the differential-coding scheme is shown in Algorithm 2.

---

**Algorithm 2:** Differential-coding Scheme

**Input:** ROI location information in one frame (assume n ROIs totally) and the information in the previous encoded frame (assume totally m ROIs)
**Output:** Bitstream containing ROI information

**if** (current frame is intra-coded frame)
  apply direct-coding scheme
**else**
  /* apply differential-coding scheme */
  /* encode old ROIs (normal + disappear) */
  $j$ = 1;
  $skip$ = 0;
  set *flag buffer* to null;      /* Store flags */
  encode $n$ into the bitstream;
  **for** $i$ = 1 to m
    $R^p_i$ = the $i$-th ROI in the sequence in the previous encoded frame, with parameters $\{label^p_i, x^p_i, y^p_i, w^p_i, h^p_i\}$;
    $R_j$ = the $j$-th ROI in the sequence in the current frame, with parameters $\{label_j, x_j, y_j, w_j, h_j\}$;
    **if** ($label_j > label^p_i$)  /* $R_j$ disappears */
      $skip = skip+1$;
      push back flag(0) into *flag buffer*;
    **else**   /* $label_j == label^p_i$ */
      $D = (x_j, y_j, w_j, h_j) - (x^p_j, y^p_j, w^p_j, h^p_j)$
      **if** ($D$ == $(0,0,0,0)$)  /* $R_j$ has same location */
        $skip = skip+1$;
        push back flag(1) into *flag buffer*;
        $j = j+1$;
      **else**
        encode $skip$, *flag buffer*, $D$ into the bitstream sequentially;
        $j = j+1$;
        $skip$ = 0;   /* initialize the parameter */
        set *flag buffer* to null;  /* initialize the parameter */
      **end**
    **end**
  **end**
  /* Before old ROIs end, write remains in the *flag buffer* */
  encode $skip$, *flag buffer* into the bitstream sequentially;
  /* encode appeared ROIs */
  **for** $i = j$ to $n$
    $R_i$ = the $i$-th ROI in the sequence in the current frame;
    **if** ($i == j$)
      encode $x_i, y_i, w_i, h_i$ into the bitstream sequentially;
    else
      $d_i = label_i - label_{i-1} - 1$;
      encode $d_i, x_i, y_i, w_i, h_i$ into the bitstream sequentially;
    **end**
  **end**
**end**

---

Several things need to be mentioned about the differential-coding scheme:

1) Since there is no reference frame for the first frame in a GOP, we apply direct-coding scheme for the first frame.

2) We define three types of ROIs: the ROI which exists in both the previous encoded frame and the current frame (normal ROI); the ROI which exists in the previous encoded frame but disappear in the current frame (disappeared ROI); and the ROI which newly appear in the current frame (appeared ROI). Some examples are shown in Fig. 3.

3) We apply different strategies on different ROI types. For a normal ROI which has the same location parameters as its corresponding one in the previous frame (i.e. $D$ in Eqn. (2) is {0, 0, 0, 0}), we allocate a skip flag to indicate such condition instead of encoding four zeros. For other normal ROIs, we encode the four differential numbers in Eqn. (2). Furthermore, for a disappearing ROI, we allocate a disappear flag to indicate such condition. Similarly, for an appeared ROI, we directly encode the four parameters of the ROIs (i.e. $x$, $y$, $w$, $h$), since no corresponding ROI in the previous encoded frame can be used as a reference.

4) In our scheme, appearing ROIs are always allocated with a new label and coded at the end of the ROI bitstream.

Fig. 3 shows an example of using the differential- coding scheme to encode ROIs in a frame.

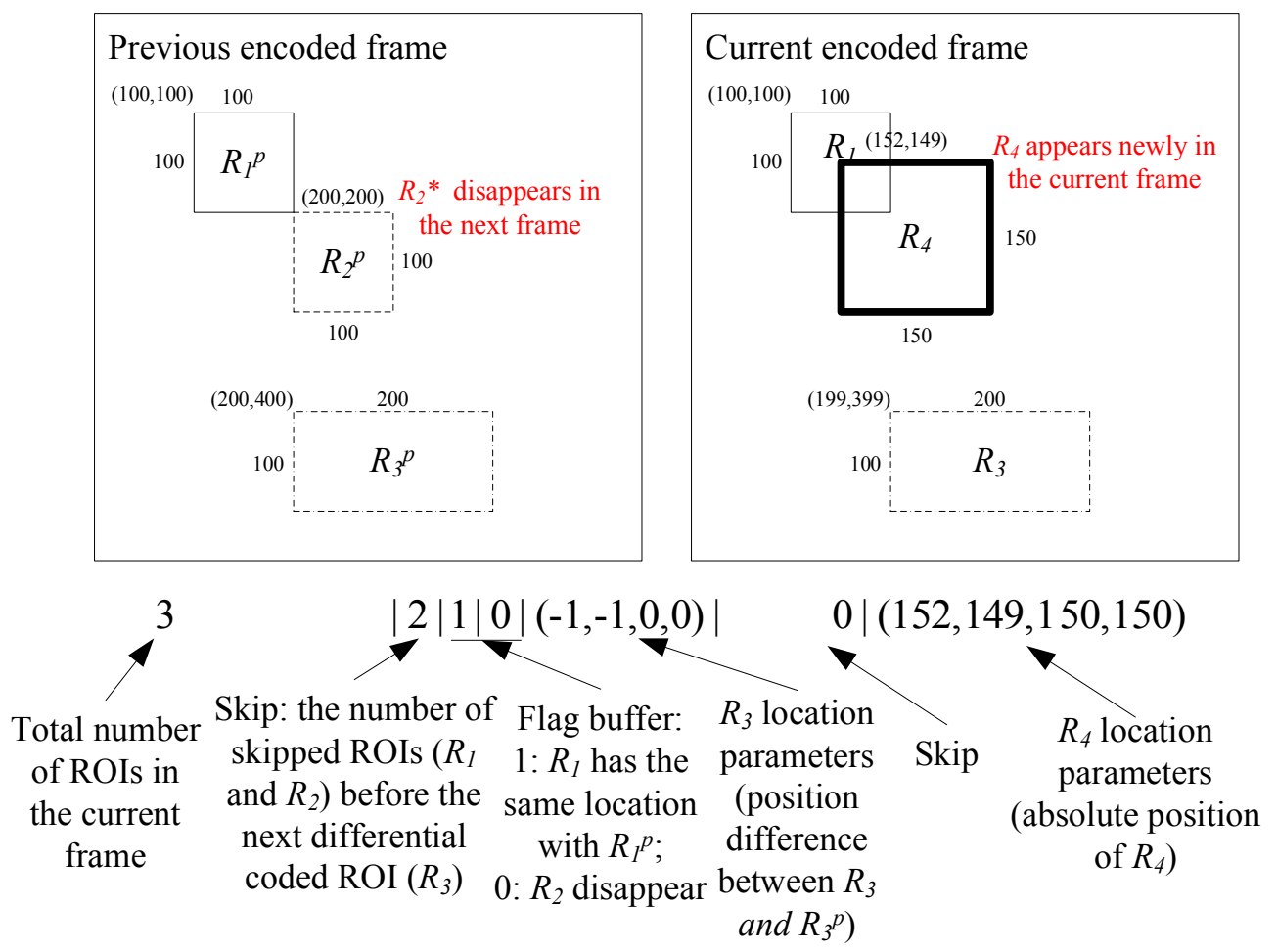

Fig. 3 An coding example of using the differential-coding scheme: assume that the label of $R_i$ is $Label_i$=$i$.

### *D. An Extended Scheme: Reconstructed-Coding Scheme*

In the differential-coding scheme, we only use the temporal correlation among ROIs for reducing coding redundancy. In practice, since many ROIs in a frame can be estimated by object detection and recognition algorithms [5], the ROI location information can be further compressed by utilizing the prediction information of these algorithms. Therefore, we also propose an extended reconstructed-coding scheme to further reduce ROI coding redundancy. Please notice that this scheme has not been included in ROI coding method in AVS2 standard.

The framework of the reconstructed-coding scheme is described by Fig. 4. More specifically, at the encoder side, after then encoding of the current frame, we apply object detection algorithms to derive predicted ROI regions from the reconstructed frame (i.e., the frame reconstructed from the current encoded frame). Then, the actual ROI locations will be differentiated with these predicted ROI locations and the location difference will be encoded and transmitted. At the decoder side, after reconstruction of the current frame, the same object detection algorithms are used to derive predicted ROI regions. These predicted ROI locations will be combined with the decoded location differences to recover the actual ROI locations.

Since the number of predicted ROIs may not be the same as the actual ones, in order to precisely recover ROI regions, we assign actual ROI with its closest predicted ROI and encode their differences. Furthermore, considering that multiple actual ROIs may correspond to one predicted ROI, we also encode the number of the actual ROIs in the bitstream. That is, we will encode in the bitstream of each predicted ROI part the number of actual ROIs associated with it, followed by the label number and location differences of each actual ROI. A coding example of reconstructed-coding scheme is shown in Fig. 5, and the above point will be illustrated visually in the meantime.

Algorithm 3 shows the detailed process of the reconstructed-coding scheme.

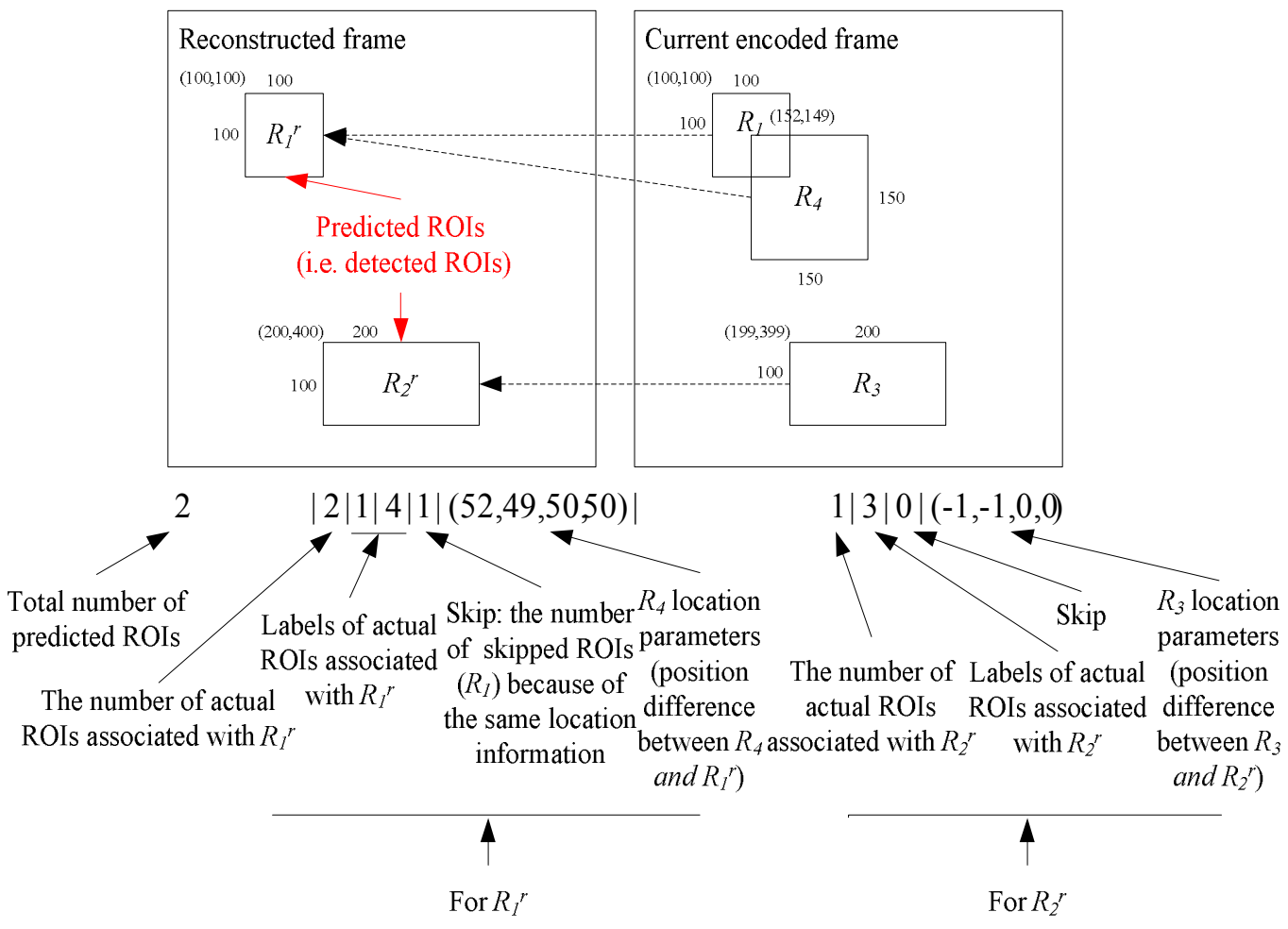

Fig. 5 An coding example of using the reconstructed-coding scheme: assume that the label of $Label_i$=$i$.

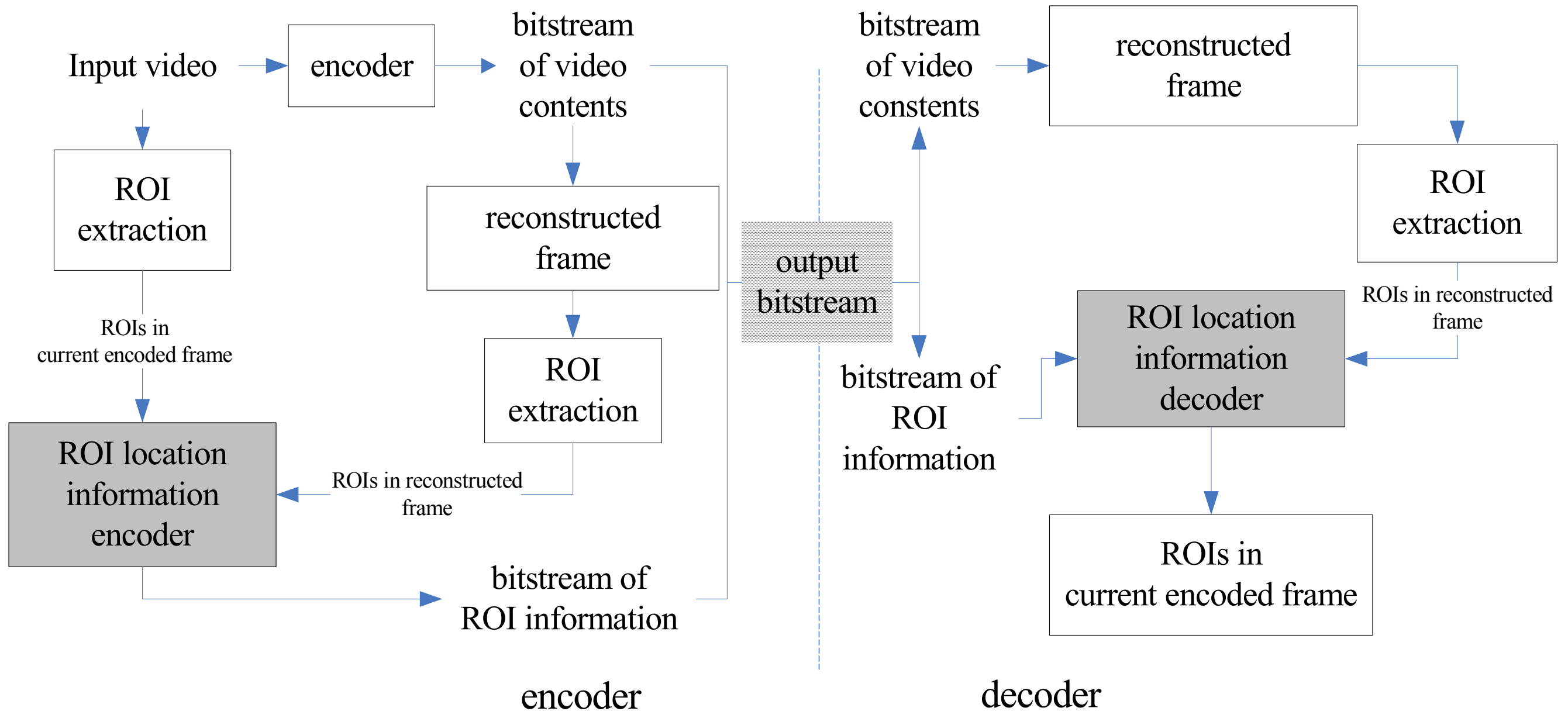

Fig. 4 The framework of Reconstructed-Coding Scheme

**Algorithm 3:** Reconstructed-coding scheme

**Input:** ROI location information in one frame (assume n ROIs totally) and the information in the reconstructed frame (assume m ROIs totally)

**Output:** Bitstream containing ROI information

encode m into the bitstream;
*skip* = 0;
**for** $i$ = 1 to m (the number of predicted ROIs)
  $R^r_i$ = the $i$-th predicted ROI in the reconstructed frame, with parameters $\{label^r_i, x^r_i, y^r_i, w^r_i, h^r_i\}$;
  $k$ = the number of actual ROIs in the current frame associated with $R^r_i$;
  encode $k$ and all the labels of the actual ROIs associated with $R^r_i$ into the bitstream sequentially;
  **foreach** actual ROIs associated with $R^r_i$ ($R_l$, $l$=1,2,…,$k$, with parameters $\{label^r_i, x^r_i, y^r_i, w^r_i, h^r_i\}$)
    $D = (x_l, y_l, w_l, h_l) - (x^r_i, y^r_i, w^r_i, h^r_i)$;
    **if** ($D$ == (0,0,0,0))   /* $R_l$ and $R^r_i$ have the same location */
      *skip* = *skip*+1;
    **else**
      encode *skip*, $D$ into the bitstream (in order);
      *skip* = 0;   /* initialize the parameter */
    **end**
  **end**
**end**

### *E. Syntax in the Bitstream*

In our tool, we use binary (0 and 1) to encode flags, use universal variable length coding (UVLC) [6] to encode difference, and use fixed length coding (FLC) label values and absolute position of ROIs. The coded bitstream is added in front of the bitstreams of a frame. Fig. 6 shows an example of the bitstream created by our differential-coding scheme.

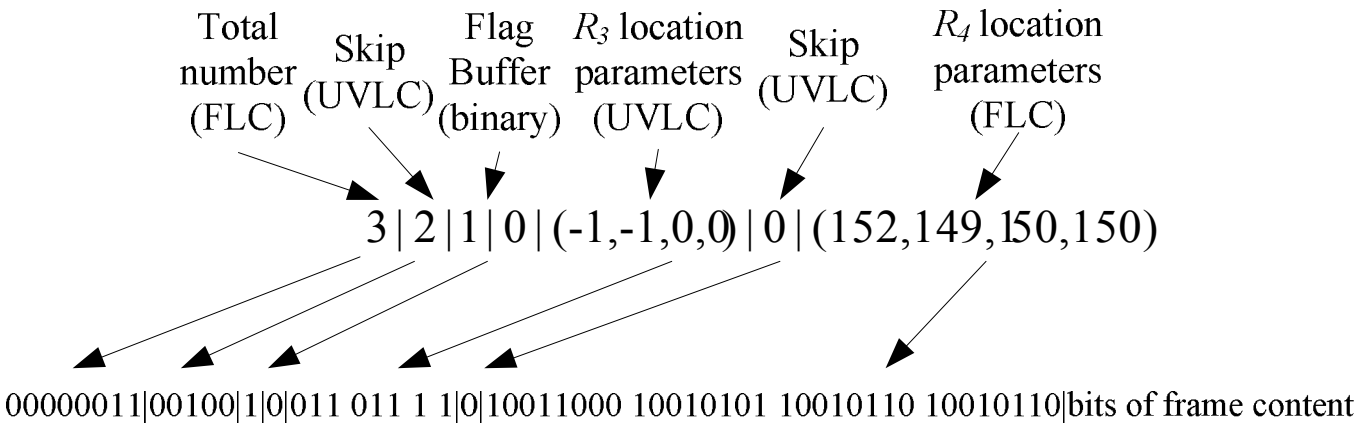

Fig. 6 An example of the bitstream for differential-coding scheme

## III. Experimental Results

In this section, we show experiments results of the three ROI coding schemes described in Section II. In the experiments, we use AVS2 RD6.1 as the test platform [7]. The picture coding structure is Hierarchical-B (one P frame with seven B). The details of test conditions can be found at [9].

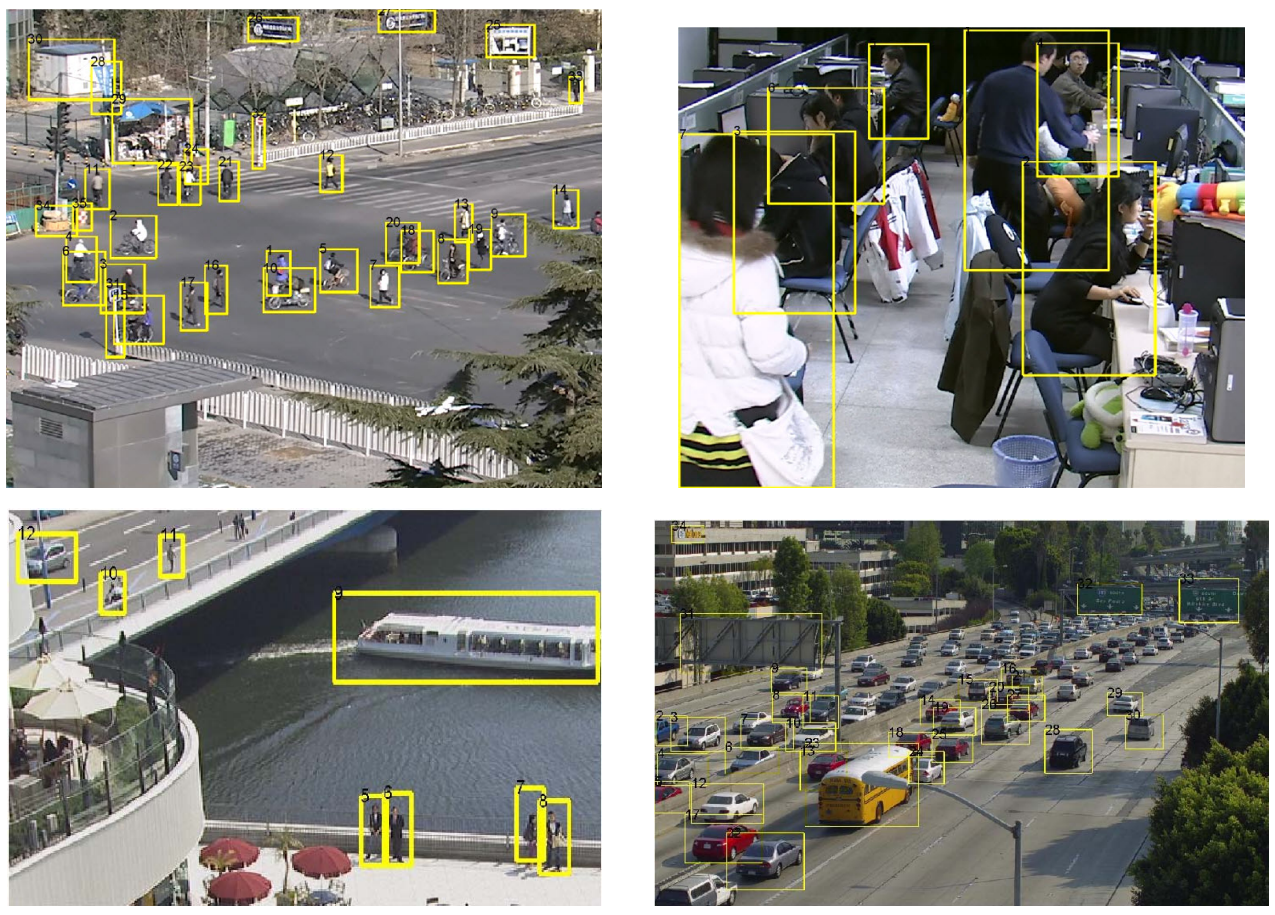

Fig. 7 Some example video frames and the ROIs of our experimental videos (standard AVS video)

We test the coding performance on AVS2 surveillance profile's common test sequences where ROIs are pre-defined for each sequence. Besides, the object detection algorithm in [5] is utilized to achieve predicted ROI regions in the reconstructed-coding scheme. Some tested video frames and the ROIs are illustrated in Fig. 7.

In the first experiment, we perform tests of the direct-coding scheme and the differential-coding scheme with different video sequences under various quantization parameters (QPs). The experiment results are shown in Table 1. Since fix quantization parameter is used in each experiment, the PSNR values are the same for different methods and thus only bit rates are listed in Table 1.

From Table 1, we can see that if we use the direct-coding scheme, the bit rate is high due to the large number of bits in coding ROI locations. The ROI overhead becomes obvious when the number of ROIs is large. Considering that large ROI number may exist frequently in many applications such as surveillance and video conference, it is desirable to reduce ROI location signaling bits. Comparatively, the bit rates by the differential method are substantially reduced by utilizing the temporal correlation among ROIs. Compared to the direct coding method, the differential method can achieve 2.5% bitrate reduction on average, and up to 6% bitrate reduction.

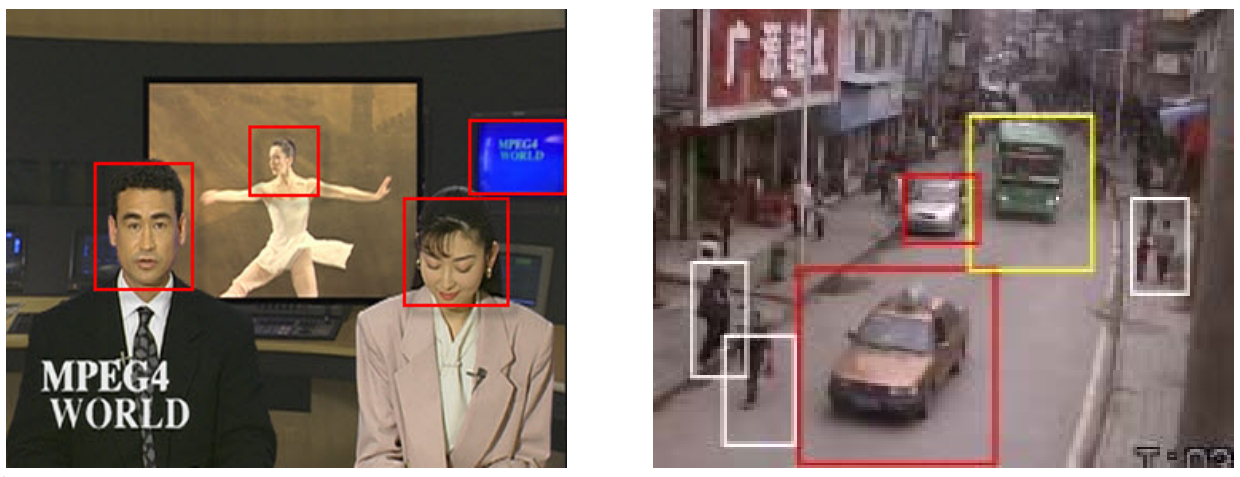

Fig. 8 Some example video frames and the ROIs in the second experiment

In the second experiment, we further perform experiments of the extended reconstructed-coding scheme. The number of encoded frames for each video sequence is 100 frames. Table 2 shows the results of the reconstructed-coding scheme on some video sequences as in Fig. 8. In order to ease comparison, we also include the corresponding results of the direct-coding and the differential-coding schemes in Table 2.

From Table 2, we can see that the reconstructed-coding scheme can also reduce bit rate by utilizing the predicted ROI regions. Moreover, comparing the reconstructed-coding and differential-coding schemes, we can see that both schemes have their own advantages. If the object detection algorithm is accurate and the predicted ROIs are close to their corresponding actual ROIs, the reconstructed-coding scheme can achieve better results than the differential-coding scheme. Otherwise, the differential-coding scheme creates better performance. Furthermore, we can also see that when the QP value increases, the performance of the reconstructed-coding schemes will be decreased since the qualities of the reconstructed frames decrease with larger QPs. However, we believe that the effect from QP parameters can be further reduced by utilizing more sophisticated detection techniques [8].

## IV. Conclusion

In this paper, we introduce a coding tool for encoding ROI locations in AVS2 standard. The coding tool has three different coding schemes which encode ROI locations directly, by using temporal information, and by using predicted ROI information from object detection algorithms. Experimental results shows that utilizing temporal and prediction information can effectively reduced the overhead for encoding ROI regions. Further work will include combining the schemes of differential-coding and reconstructed-coding such that the coding performance can be further improved.

Table 1 Experiments on the schemes in the coding tool

| seq. | encoded frames | resolution | ROI/frame | QP | Bit Rate (kbps) @ 30Hz | | |
|---|---|---|---|---|---|---|---|
| | | | | | Original | Direct-coding | Differential-coding |
| BQquare | 600 | 416x240 | 2~8 | 31 | 1049.32 | 1061.72(+1.18%) | 1051.94(+0.25%) |
| | | | | 38 | 503.74 | 516.13(+2.46%) | 506.36(+0.52%) |
| | | | | 45 | 249.81 | 262.20(+4.96%) | 252.43(+1.05%) |
| | | | | 51 | 122.65 | 135.05(+10.11%) | 125.28(+2.14%) |
| Traffic | 150 | 2560x1600 | 30~40 | 31 | 16078.21 | 16086.16(+0.05%) | 16079.52(+0.01%) |
| | | | | 38 | 10184.15 | 10192.10(+0.08%) | 10185.46(+0.01%) |
| | | | | 45 | 5366.03 | 5373.98(+0.15%) | 5367.35(+0.02%) |
| | | | | 51 | 2300.17 | 2308.12(+0.35%) | 2301.48(+0.06%) |
| Crossroad | 600 | 720x576 | 30~40 | 31 | 2470.55 | 2510.48(+1.62%) | 2478.05(+0.30%) |
| | | | | 38 | 1151.40 | 1191.33(+3.47%) | 1158.90(+0.65%) |
| | | | | 45 | 503.57 | 543.49(+7.93%) | 511.06(+1.49%) |
| | | | | 51 | 228.52 | 268.45(+17.47%) | 236.02(+3.28%) |
| Office | 600 | 720x576 | 7~15 | 31 | 1707.29 | 1722.68(+0.90%) | 1711.04(+0.22%) |
| | | | | 38 | 869.59 | 884.99(+1.77%) | 873.35(+0.43%) |
| | | | | 45 | 441.44 | 456.84(+3.49%) | 445.20(+0.85%) |
| | | | | 51 | 222.35 | 237.74(+6.92%) | 226.10(+1.69%) |
| Intersection | 150 | 1600x1200 | 35~50 | 31 | 8346.84 | 8396.38(+0.59%) | 8359.44(+0.15%) |
| | | | | 38 | 4429.66 | 4479.21(+1.12%) | 4442.27(+0.28%) |
| | | | | 45 | 2267.58 | 2317.12(+2.19%) | 2280.18(+0.56%) |
| | | | | 51 | 1096.99 | 1146.53(+4.52%) | 1109.59(+1.15%) |

Table 2 Experiments on three optional schemes
(We encode 100 frames for each video sequence)

| seq. | QP | resolution | ROI/frame | Bit Rate (kbps) @ 30Hz | | | |
|---|---|---|---|---|---|---|---|
| | | | | Original | Direct-coding | Differential-coding | Reconstructed-coding |
| I | 16 | 176×144 | 4~6 | 905.48 | 911.37(+0.65%) | 908.11(+0.29%) | 907.74(+0.25%) |
| I | 28 | 176×144 | 4~6 | 130.02 | 136.01(+4.61%) | 132.53(+1.93%) | 132.43(+1.85%) |
| I | 36 | 176×144 | 4~6 | 35.79 | 40.38(+12.82%) | 38.16(+6.63%) | 38.37(+7.22%) |
| II | 16 | 352×288 | 5~7 | 817.25 | 823.46(+0.76%) | 821.09(+0.47%) | 821.50(+0.52%) |
| II | 28 | 352×288 | 5~7 | 111.32 | 118.30(+6.27%) | 114.31(+2.69%) | 114.38(+2.75%) |
| II | 36 | 352×288 | 5~7 | 37.62 | 42.84(+13.88%) | 40.72(+8.25%) | 40.89(+8.68%) |
| III | 28 | 176×144 | 5~7 | 103.86 | 110.44(+6.34%) | 106.50(+2.54%) | 106.33(+2.38%) |
| IV | 28 | 864×480 | 14~17 | 142.23 | 162.68(+14.38%) | 152.43(+7.17%) | 153.07(+7.62%) |
| V | 28 | 640×480 | 27~30 | 131.77 | 166.19(+26.12%) | 147.93(+12.26%) | 147.44(+11.89%) |
| VI | 28 | 640×480 | 55~63 | 265.54 | 331.29(+24.76%) | 293.34(+10.47%) | 295.31(+11.21%) |

**Acknowledgements**

This paper is supported in part by the following grants: National Science Foundation of China (No. 61001146, 61025005 and 61102100), the Chinese national 973 project (2010CB731401), the SMC grant of SJTU, and Shanghai Pujiang Program (12PJ1404300).